\documentclass[%
 reprint,
 showpacs,preprintnumbers,
 amsmath,amssymb,
 aps,
 prb,
]{revtex4-1}
\usepackage{dcolumn}
\usepackage{subfigure}
\usepackage{graphicx,color}
\usepackage{mathrsfs}

\makeatletter
\def\btt#1{\texttt{\@backslashchar#1}}
\DeclareRobustCommand\bblash{\btt{\@backslashchar}} \makeatother

\begin{document}

\title{Dynamical conductivity in topological nodal-line semimetal ZrSiS}

\author{Tetsuro Habe and Mikito Koshino}
\affiliation{Department of Physics, Osaka University, Toyonaka, Osaka 560-0043, Japan}

\date{\today}

\begin{abstract}
ZrSiS is one of the strong candidates for realistic nodal-line semimetal.
We theoretically investigate the dynamical conductivity in ZrSiS by using a multi-orbital theoretical model based on the first-principles band calculation.
We find that the dynamical conductivity in the clean limit is actually not frequency independent unlike the ideal Dirac model, while nearly flat dependence is achieved by introducing the energy broadening possibly induced by the disorder.
The results can be applied to other compounds with the similar crystal structure, such as ZrSiSe, ZrSiTe, and HfSiS.
\end{abstract}

\maketitle
\section{Introduction}
Nodal semimetal is a novel class of topological material and attracts much attention in condensed matter physics\cite{Murakami2007}.
It is characterized by the gapless energy band structure, where the conduction and valence band touch with each other, and the band structure linearly disperses in the momentum space around the band touching point (node).
The Dirac and Weyl semimetals\cite{Wan2011,Yang2011,Burkov2011-weyl,Borisenko2013,Neupane2013} are three-dimensional (3D) point-node semimetal, where the band touching occurs at discrete points in the 3D Brillouin zone.
On the other hand, the nodal-line semimetal\cite{Burkov2011,Habe2014,Phillips2014,Fang2015} is another type of nodal semimetal in which the energy bands stick on a line in the reciprocal space.
Recently, several theoretical works predicted 3D nodal-line semimetals in condensed matter systems.\cite{Burkov2011,Habe2014,Phillips2014,Fang2015,Lilia2015,Yu2015,Yamakage2016}, and various unusual phenomena related to the nodal-line were also proposed\cite{Ramamurthy2017,Carbotte2016,Koshino2016,Mukherjee2017,Syzranov2017,Mikitik2018,Rui2018}.

Experimentally, the evidence of the nodal-line feature has been detected in several materials by angle resolved photoemission spectroscopy\cite{Wu2016,Takane2016,Chen2017} and magnetotransport measurements\cite{Emmanouilidou2017,Kumar2017}.
In particular, a series of compounds, ZrSiS, ZrSiSe, ZrSiTe, and HfSiS, are predicted to be nodal-line semimetals with the similar electronic structure, and the characteristics of nodal-line was actually confirmed in recent experiments.\cite{Schoop2016,Neupane2016,Lv2016,Lou2016,Topp2016,Hosen2017,Zhang2017}
Here the band crossing appears near the Fermi energy, and therefore it serves an ideal platform for experimental exploration of the nodal-line semimetals\cite{Lv2016,Ali2016,Emmanouilidou2017,Kumar2017,Matusiak2017,Zhang2017,Lodge2017}.
Recently, the optical absorption of ZrSiS was measured in a broad frequency range,\cite{Schilling2017} and it was shown that the absorption ratio is nearly independent of photon energy up to $\sim350$ meV.
Such a flat frequency dependence may seem a characteristic property of the ideal linear dispersion such as graphene\cite{Carbotte2016,Mukherjee2017,Ahn2017,Barati2017}.
In ZrSiS, however, the linear approximation is valid only in relatively narrower energy region less than 100 meV.
There must be an alternative mechanism to realize the flat dynamical conductivity up to 350 meV, under the realistic complex electronic structure.

In this paper, we calculate the dynamical conductivity of ZrSiS using a multi-orbital tight-binding model based on the first-principles band calculation. 
We discuss about the frequency dependence in relation to the electronic structure.
We find that the dynamical conductivity in the clean limit is actually not frequency independent, while nearly flat dependence is achieved by introducing the energy broadening, which is possibly induced by the disorder.
Our study focuses on ZrSiS but the results can be applied to other compounds with the similar crystal structure, e.g., HfSiS.

This paper is organized as follows.
First, we calculate the band structure in ZrSiS by a first-principles method, and we investigate the electronic structure and nodal-lines in detail in Sec.\ \ref{Sec_first-principles}.
In Sec.\ \ref{Sec_dynamical_conductivity}, we introduce the multi-orbital tight-binding model, and numerically calculate the dynamical conductivity with various polarization directions.
Furthermore, we analyze the relation between the dynamical conductivity and the electronic structure by using a simple model in Sec.\ \ref{Sec_analysis}.
The conclusion is given in Sec.\ \ref{Sec_conclusion}.

\begin{figure}[htbp]
\begin{center}
 \includegraphics[width=70mm]{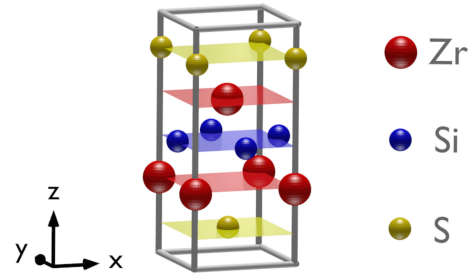}
\caption{Atomic structure in the unit cell of ZrSiS. 
 }\label{Unit_cell}
\end{center}
\end{figure}
\setcounter{figure}{2}
\begin{figure*}[tbp]
\begin{center}
\includegraphics[width=180mm]{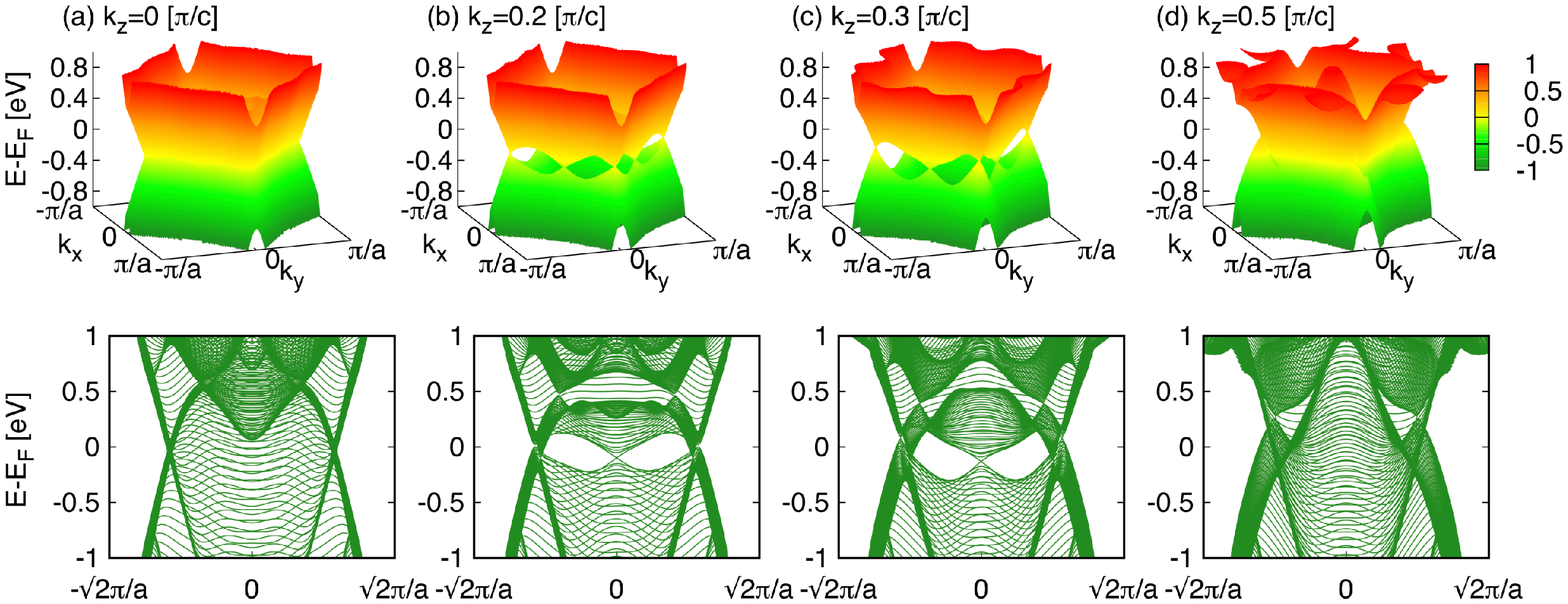}
\caption{
(Upper panels) First-principles band structure of ZrSiS in the low-energy region, plotted in the two-dimensional momentum space $(k_x, k_y)$ at several fixed values of $k_z$. (Lower panels) The same energy bands projected to $(k_x, k_y)\parallel (1,1)$ direction. 
 }\label{3D_bands}
\end{center}
\end{figure*}
\setcounter{figure}{1}
\begin{figure}[htbp]
\begin{center}
 \includegraphics[width=85mm]{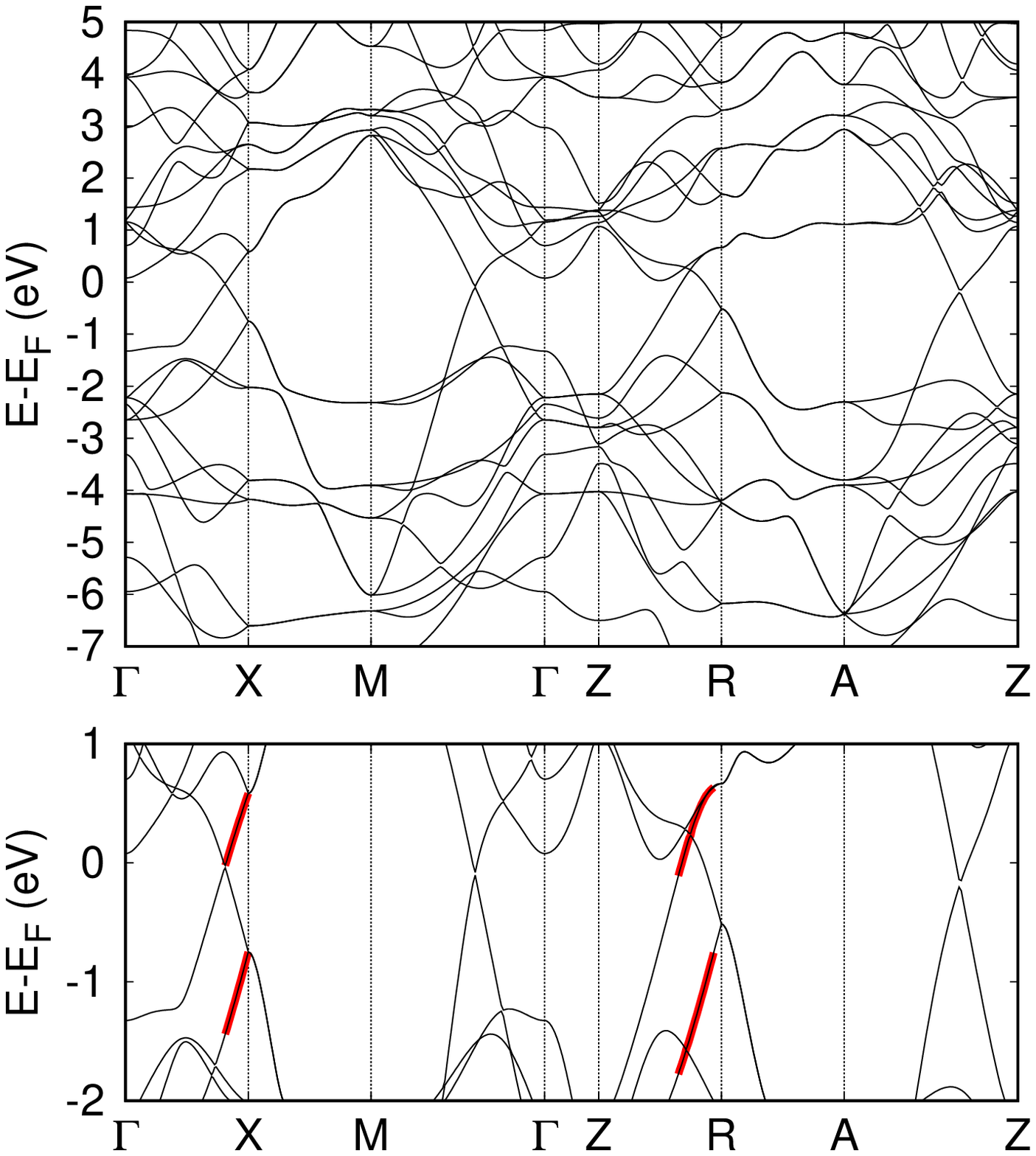}
\caption{(Top) First-principles band structure of ZrSiS. (Bottom) Zoom-in plot of the low-energy region. The red lines indicate the band states associated with the optical transitions at $\omega\sim1.3$ eV (see the text).
 }\label{Band_structure}
\end{center}
\end{figure}
\section{First-principles band calculation}\label{Sec_first-principles}
ZrSiS is a tetragonal crystal, and its unit cell contains two Zr atoms, two Si atoms, and two S atoms as shown in Fig.\ \ref{Unit_cell}.
We set the $z$ axis to be parallel to the $c$ axis, and the $x$ and $y$ axes as shown in the figure.
We calculate the band structure by the first-principles calculation code, quantum ESPRESSO\cite{quantum-espresso}.
Here, we adopt the lattice structure in Ref.\ \onlinecite{Tremel1987,Wang1995}, where the lattice constants are $a=3.545$ \AA\  and $c/a=2.273$, and the atomic positions are given in Table.\ \ref{atomic_positions}.
\begin{table}
\begin{center}
\caption{Relative atomic positions in a unit cell of ZrSiS, in units of the lattice constant.}
\begin{tabular}{c c c c c }
\hline
\hline
&x [a]&y [a]&z [c]\\ \hline
S$^{(1)}$& 0.50 & 0.50 & 0.12\\
Zr$^{(1)}$& 0.00 & 0.00 & 0.23\\
Si$^{(1)}$& 0.00 & 0.50 & 0.50\\
Si$^{(2)}$& 0.50 & 0.00 & 0.50\\
Zr$^{(2)}$& 0.50 & 0.50 & 0.77\\
S$^{(2)}$& 0.00 & 0.00 & 0.88\\ \hline \hline
\end{tabular}\label{atomic_positions}
\end{center}
\end{table}
In this calculation, we employ a PAW type pseudopotential with GGA functional, the cut-off energy of plane wave basis 150 Ry, and the conversion criterion 10$^{-8}$ Ry. 
We omit the spin-orbit coupling, which opens a small gap $\sim10$ meV, in the band structure but the effect of this coupling appears below 100 K in the low photon energy region $\omega<20$ meV\cite{Schilling2017}.

We show the calculated band structure in Fig.\ \ref{Band_structure}. 
The conduction and valence bands are touching at off-symmetric points in the region $\Gamma$-$X$, $\Gamma$-$M$, $Z$-$R$, and $Z$-$A$, and the energy bands show linear dispersion around the nodal points.
To see the low energy band structure in more detail, we plot the energy band on the two-dimensional momentum space ($k_x, k_y$) at fixed $k_z$'s in Fig.\ \ref{3D_bands} (a) to (d). 
The lower figure in each panel is the side view from ($k_x, k_y$)$\parallel$(1,1).

In $0< k_z<\pi/(2c)$ [Figs.\ \ref{3D_bands}(b) and (c)], the conduction band and valence band touch at eight points arranged in a diamond-like shape on $k_xk_y$-plane.
By changing $k_z$, the touching points continuously move on $k_xk_y$-plane, forming nodal lines along $k_z$ direction.
At $k_z=0$ [Fig.\ \ref{3D_bands}(a)] and $\pi/2c$ [Fig.\ \ref{3D_bands}], the band touching occurs entirely on the diamond, and this gives another nodal line parallel to $k_xk_y$-plane.
\begin{figure}[htbp]
\begin{center}
 \includegraphics[width=85mm]{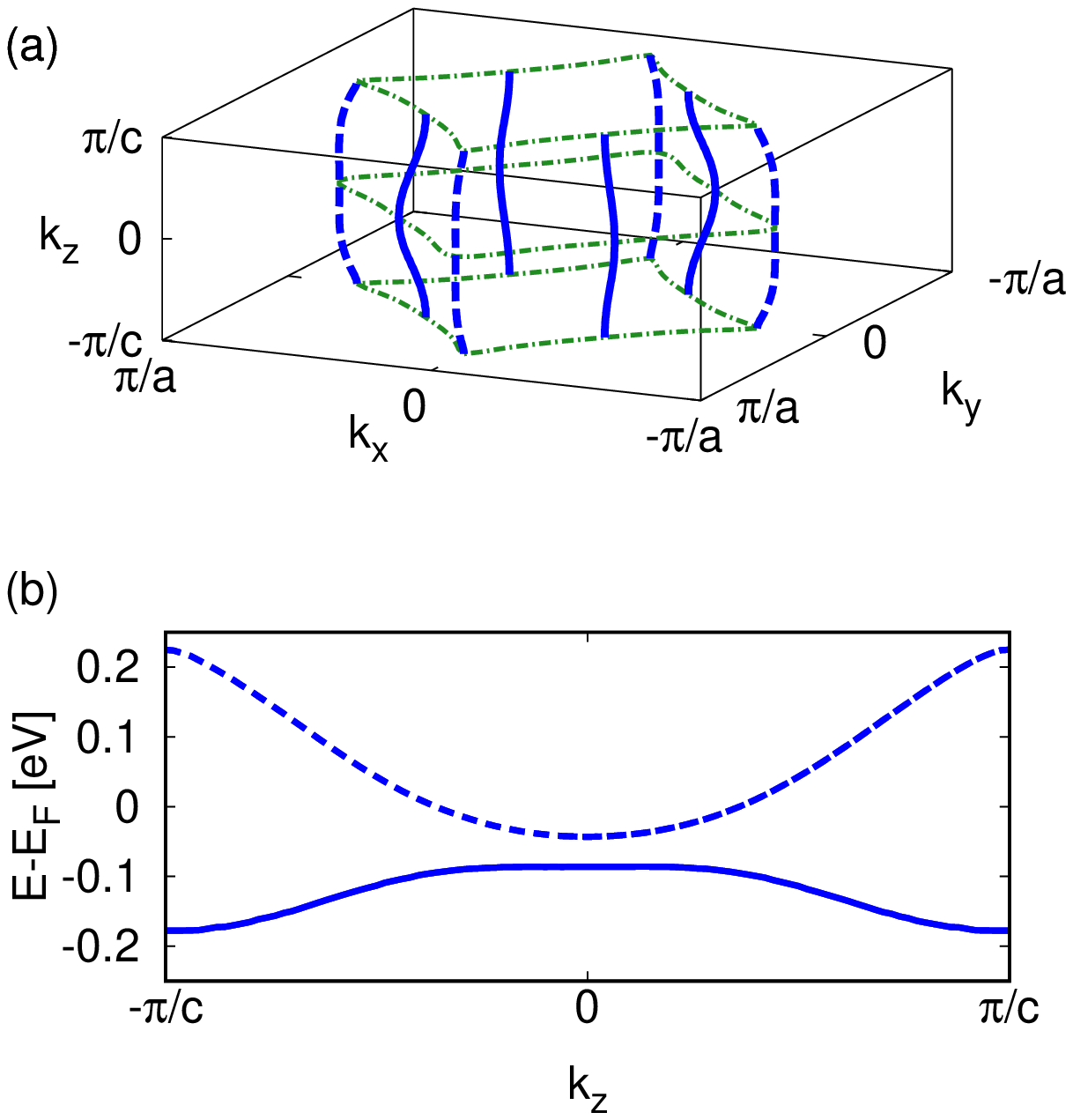}
\caption{
(a) Network structure of nodal-lines in the first Brillouin zone of ZrSiS. (b) Energy dispersion of the nodal-lines along $k_z$. The dashed and solid nodal curves correspond to those in (a).
 }\label{Nodal-line}
\end{center}
\end{figure}
The overall structure of the nodal-lines in three-dimensional Brillouin zone is illustrated in Fig.\ \ref{Nodal-line}(a), where the horizontal and vertical nodal-lines form a cage like structure.
The top and bottom of the cage connect to itself at the Brillouin zone boundary, $k_z=\pm\pi/c$.
The nodal-lines on $k_xk_y$-plane are located at a single energy, while the one along $k_z$ direction has the energy dispersion as shown in Fig.\ \ref{Nodal-line}(b).

\section{Dynamical conductivity}\label{Sec_dynamical_conductivity}
We investigate the dynamical conductivity in ZrSiS by using a multi-orbital tight-binding model based on the DFT band calculation.
Here we obtain maximally-localized Wannier functions and hopping matrix elements by Wannier90\cite{Wannier90}, to replicate the DFT band structure in Fig.\ \ref{Band_structure}.
We take five $d$-orbitals in Zr atom, $s$-orbital and three $p$-orbitals in Si and S atoms.
The tight-binding Hamiltonian $H$ is written in terms of the hopping integrals between the Wannier orbitals $|\mathrm{Zr}^{(j)},d_\mu\rangle$, $|\mathrm{Si}^{(j)},s\rangle$, $|\mathrm{Si}^{(j)},p_\nu\rangle$, $|\mathrm{S}^{(j)},s\rangle$, and $|\mathrm{S}^{(j)},p_\nu\rangle$ where $\mu=\{3z^2-r^2$, $zx$, $zy$, $x^2-y^2$, $xy\}$ and $\nu=\{x$, $y$, $z\}$, and $j=1$ and 2 represents two atomic positions in the unit cell as shown in Table.\ \ref{atomic_positions}.
We obtain the Bloch wave function of the system by diagonalizing the tight-binding Hamiltonian.

\begin{figure}[htbp]
\begin{center}
 \includegraphics[width=87mm]{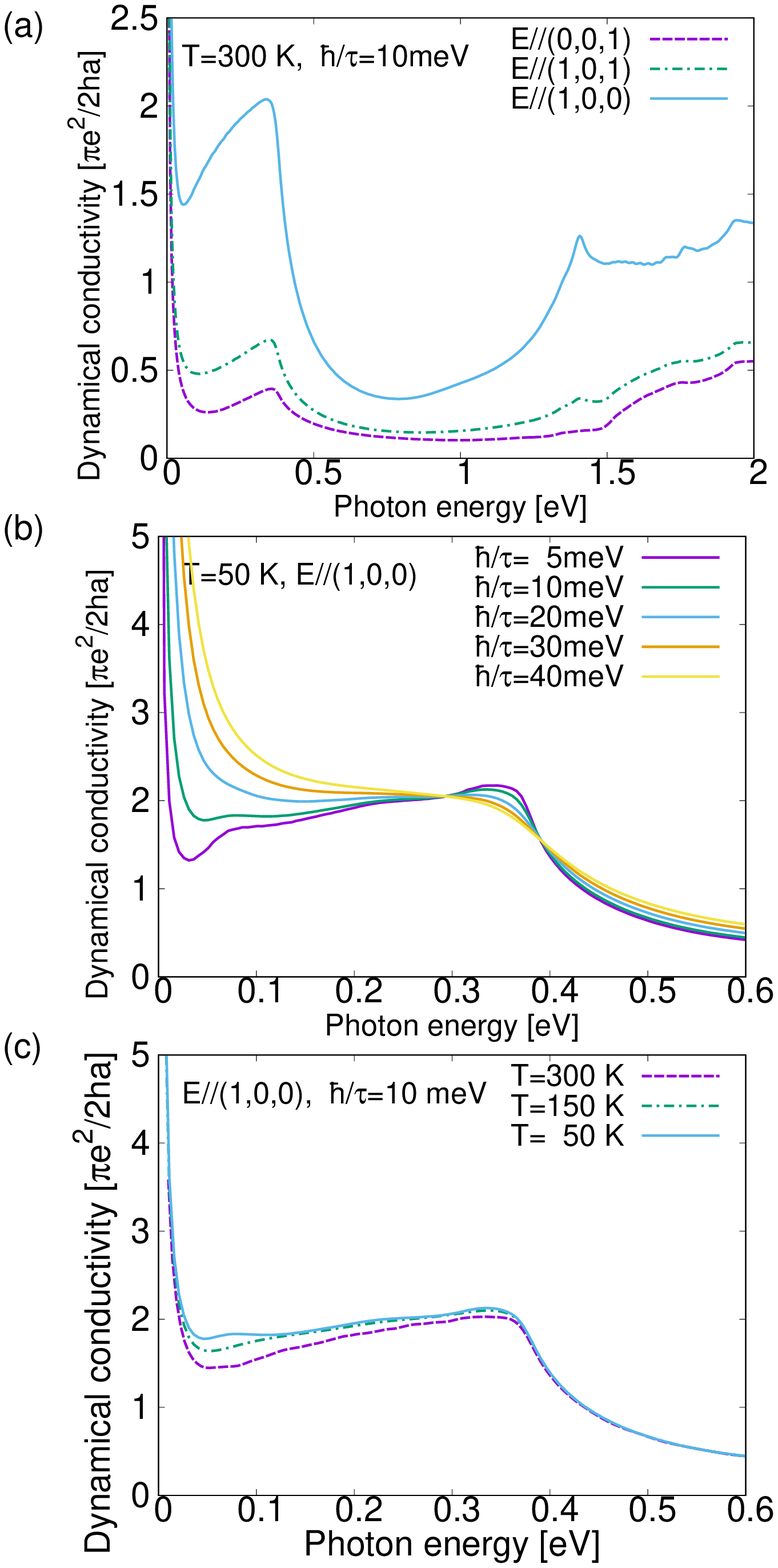}
\caption{
(a) Dynamical conductivity plotted as a function of the photon energy, in different polarization directions. (b) Similar plots for different relaxation times, and (c) for different temperatures.
 }\label{optical_conductivity}
\end{center}
\end{figure}
In the linear response theory, the dynamical conductivity $\sigma(\omega,T)$ is expressed as
\begin{align}
\sigma(\omega,T)=&\frac{2i}{\omega}\frac{e^2}{\hbar}\sum_{m<n}\int_{\mathrm{BZ}}\frac{d^3\boldsymbol{k}}{(2\pi)^3}\frac{|\langle n\boldsymbol{k}|v_\alpha|m\boldsymbol{k}\rangle|^2}{\omega+E_{m\boldsymbol{k}}-E_{n\boldsymbol{k}}+i\hbar/\tau}\notag\\
&\times(n_F(E_{m\boldsymbol{k}},T)-n_F(E_{n\boldsymbol{k}},T))
.\label{eq_optical_conductivity}
\end{align}
The optical absorption of the linear polarized light is proportional to $\sigma_1(\omega)=\mathrm{Re}[\sigma(\omega)]$. 
Here, $v_\alpha=(1/i\hbar)[x_\alpha,H]$ is a velocity operator with $\alpha$ parallel to the electric field, $|m\boldsymbol{k}\rangle$ represents the electronic eigenstate in the absence of external fields, $\tau$ is phenomenological relaxation time, and $n_F(E,T)=(1+\exp[-(E-\mu)/k_BT])^{-1}$ is the Fermi distribution function with the chemical potential $\mu$.

In Fig.\ \ref{optical_conductivity} (a), we plot $\sigma_1(\omega)$ for a linear polarized photon with polarizing directions (0,0,1), (1,0,1), and (1,0,0).
In every case, $\sigma_1(\omega)$ has the Drude peak at $\omega=0$, and after that $\sigma_1(\omega)$ monotonically increase up to $\omega\sim350$ meV and then it sharply falls.
In even higher frequencies than 1 eV, $\sigma_1(\omega)$ rises again.
The low-frequency dependence of $\sigma_1(\omega)$ is quite different from that of simple Dirac model (e.g., graphene) where the dynamical sheet conductivity is independent of $\omega$.\cite{Ando2002,Gusynin2006,Koshino2008,Stauber2008,Mak2008}
In Fig.\ \ref{optical_conductivity} (b), we show the relaxation-time dependence of dynamical conductivity at $T=50$K.
In increasing $\hbar/\tau$, the upslope in $\omega<400$ meV is gradually flattened by the spectral broadening, and almost flat $\omega$-dependence is achieved at $\hbar/\tau=30$ meV.
A similar flat dependence on frequency was actually observed in the same photon energy region in the recent light-absorption measurement. [\onlinecite{Schilling2017}]

In Fig.\ \ref{optical_conductivity} (c), we plot the dynamical conductivity of $\hbar/\tau=10$ meV at different temperatures.
In increasing temperature, we see that the conductivity slightly decrease mainly in the low frequency region.
This is because thermally-excited electrons and holes disturb the optical transition from the valence band to the conduction band.
As a result, the overall slope of the flat region slightly increases in increasing temperature. 
In larger $\hbar/\tau$, the temperature dependence becomes even smaller.

Lastly, we notice the calculated dynamical conductivity exhibits a sharp increase above a photon energy $\omega\sim1.3$ eV, which is also in a good agreement with the experimental observation [\onlinecite{Schilling2017}].
This feature is associated with the excitation between the parallel slopes of the lowest conduction and the second highest valence band indicated in the lower panel of Fig.\ \ref{Band_structure}. 
There the slopes disperse in the ($k_x,k_y$) direction, and thus the sharp peak is absent in the polarizing direction of $E//e_z$, as shown in Fig.\ \ref{optical_conductivity}.

\section{Analysis by simple model}\label{Sec_analysis}
\begin{figure}[htbp]
\begin{center}
 \includegraphics[width=80mm]{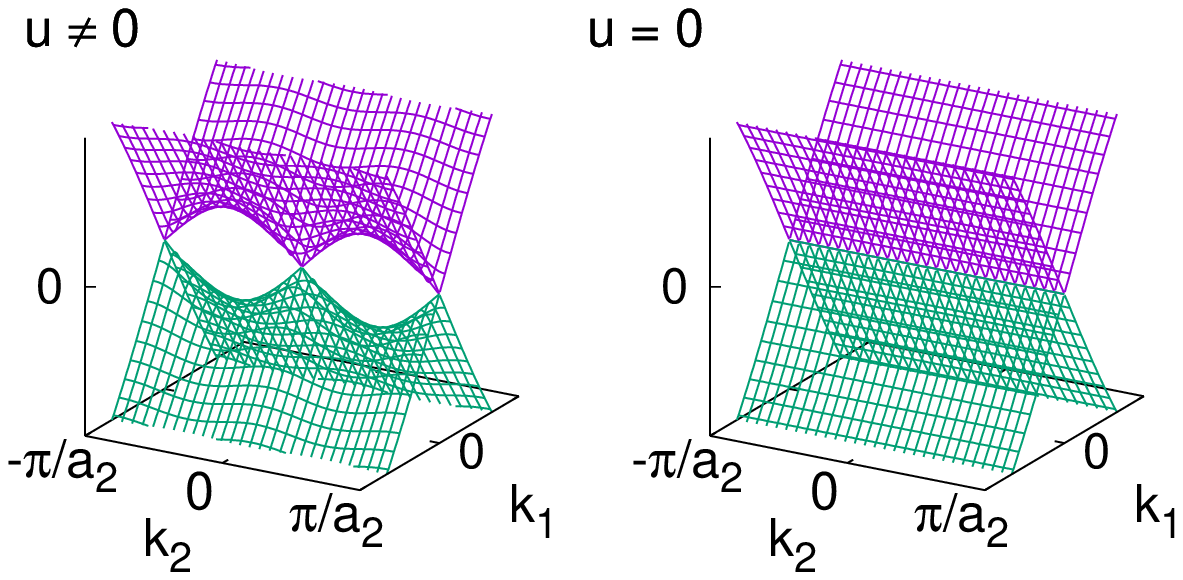}
\caption{
Band structure described by the effective Hamiltonian of Eq.\ (\ref{effective_Hamiltonian}).
 }\label{effective_bands}
\end{center}
\end{figure}

In this section, we explain the characteristic peak structure of $\sigma_1(\omega)$ at $\omega\sim350$ meV using a simple 2$\times$2 toy model.
The band structure in Fig.\ \ref{3D_bands}(a)-(d) has a diamond-like structure with four-fold rotation symmetry, and the band touching occurs at every corner and every midpoint of the diamond. 
At $k_z=0$ and $\pi/c$, the band gap closes entirely on the sides of the diamond.
We pick up the band structure on a single side of the diamond, and model it with a $2\times2$ Hamiltonian,
\begin{align}
H_{\mathrm{eff}}={v} k_1\sigma_z+{u}\sin k_2a_2\sigma_x,\label{effective_Hamiltonian}
\end{align}
where $\sigma_\nu$ is the Pauli matrix, $a_2$ is $\sim 3.545$ \AA, and $v$ is $\sim1.5\times10^{5}$ m/s for ZrSiS.
The energy dispersion of Eq.\ \ref{effective_Hamiltonian} is shown in Fig.\ \ref{effective_bands}.
The dependence of the band structure on $k_z$ can be described by the parameter $u$, and the gap-closing structure at $k_z=0$ and $\pi/c$ correspond to $u=0$.
Approximately, we have $u=u_0\sin k_z$ with $u_0\sim 170$ meV.
The parameters satisfy the condition $ua_2\ll v$, i.e., the typical velocity in the $k_1$ direction is much greater than that in the $k_2$ direction.

\begin{figure}[htbp]
\begin{center}
 \includegraphics[width=80mm]{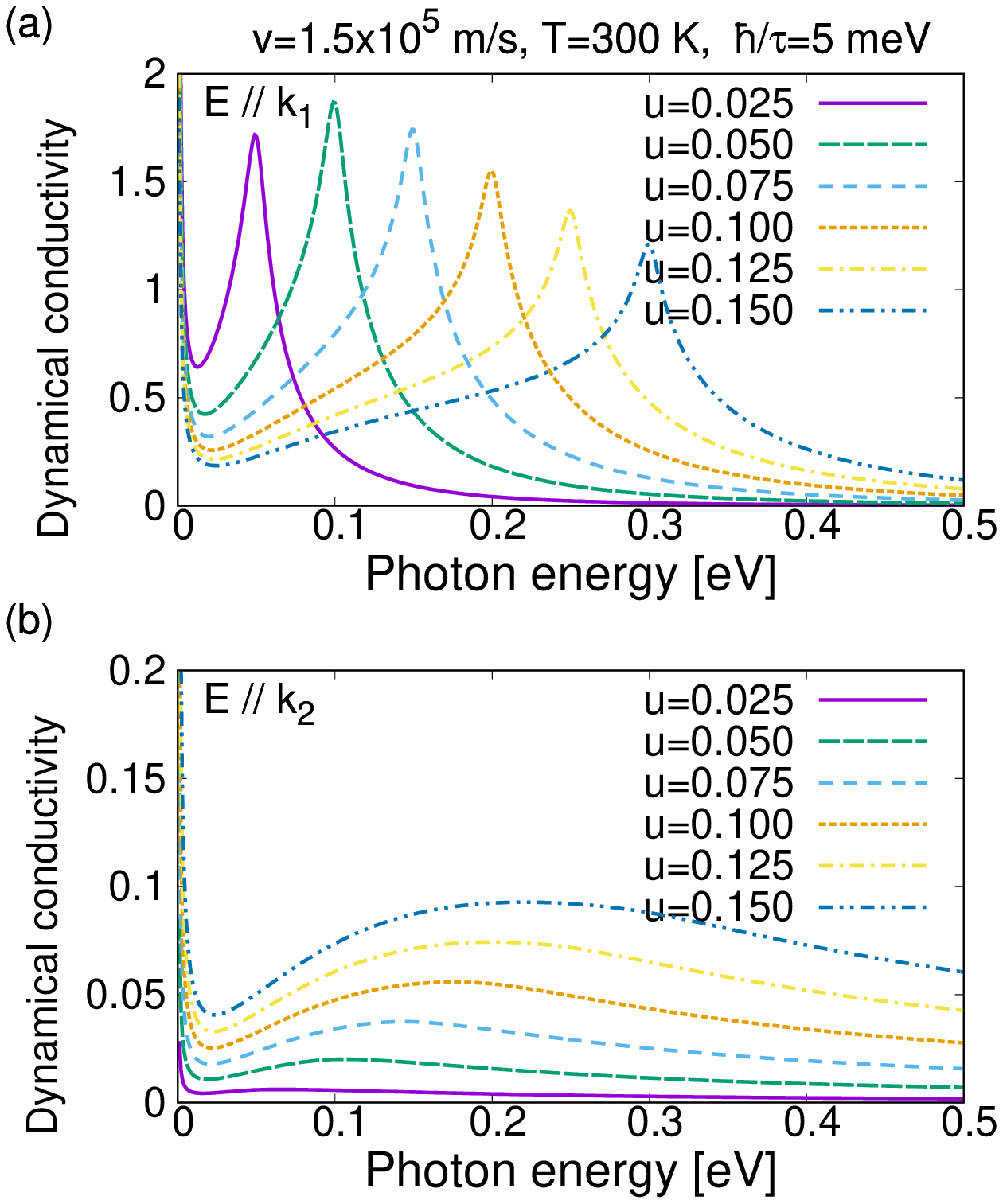}
\caption{
Dynamical conductivity as a function of photon energy, in the effective model Eq.\ (\ref{effective_Hamiltonian}) with the polarization parallel to (a) $k_1$ and (b) $k_2$. The conductivity is plotted in units of $\pi e^2/h$.
 }\label{fig_simplified_OC}
\end{center}
\end{figure}

In Fig.\ \ref{fig_simplified_OC}, we plot the two-dimensional (2D) dynamical conductivity for several $u$'s, where the photon polarizing direction is parallel to $k_1$ and $k_2$ in (a) and (b), respectively.
The numerical result exhibits a strong anisotropy where the photon polarized in the $k_1$ direction leads to much larger conductivity than that in the $k_2$ direction.
Moreover, the dynamical conductivity in (a) has a peak corresponding to the photon energy equal to the maximum split of $2u$ and then sharply falls.

The sharp drop at $\omega\sim2u$ can be explained by considering the off-diagonal velocity component between the electronic states $|+\boldsymbol{k}\rangle$ in the conduction band and $|-\boldsymbol{k}\rangle$ in the valence band:
\begin{align}
\langle+\boldsymbol{k}|\hat{v}_1|-\boldsymbol{k}\rangle
=&\frac{vu\sin k_2a_2}{\sqrt{v^2k_1^2+u^2\sin^2k_2a_2}},\label{eq_v1}\\
\langle+\boldsymbol{k}|\hat{v}_2|-\boldsymbol{k}\rangle
=&\frac{vu{k_1}a_2\cos k_2a_2}{\sqrt{v^2 {k_1}^2+u^2\sin^2k_2a_2}},\label{eq_v2}
\end{align}
where the energy split between these states is given by $\Delta E=2\sqrt{v^2 {k_1}^2+u^2\sin^2k_2a_2}$.
The mixing between the conduction band and valence band is significant when $vk_1\sim u$, where the matrix element Eq.\ (\ref{eq_v1}) is of the order of $v$.
In the opposite limit $vk_1\gg u$, Eq.\ (\ref{eq_v1}) rapidly shrinks in proportion to $\sim1/k_1$ and this leads to a sudden fall in Fig.\ \ref{fig_simplified_OC} (a).
On the other hand, the matrix element of $v_2$ in Eq.\ (\ref{eq_v2}) is always of the order of $ua_2$, and therefore it is much smaller than Eq.\ (\ref{eq_v1}), and also the peak structure does not appear.
The total dynamical conductivity in ZrSiS can be obtained as a superposition of the 2D dynamical conductivity for the different $u$'s.
The sharp fall at $\omega\sim350$ meV in Fig.\ \ref{optical_conductivity} corresponds to the maximum of $2u$.

\begin{figure}[htbp]
\begin{center}
 \includegraphics[width=80mm]{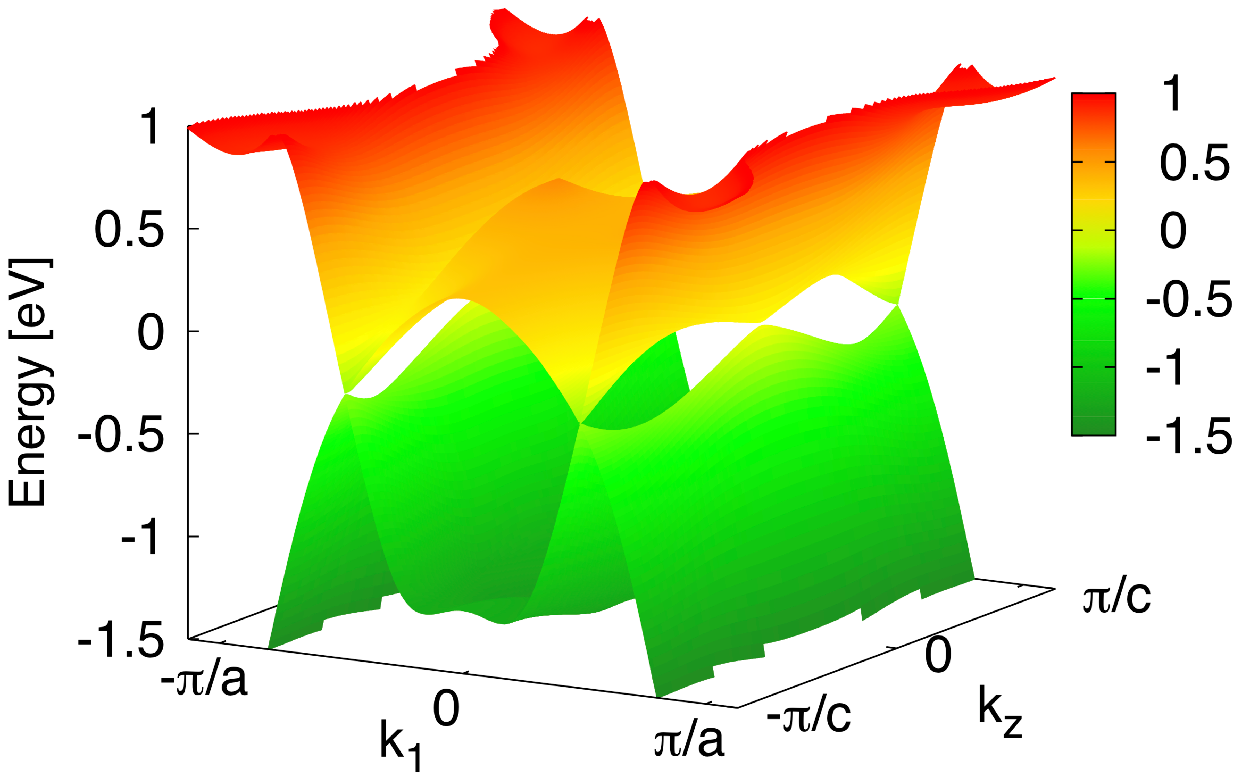}
\caption{
Band structure of ZrSiS in the 2D momentum space $k_z$ and $k_1\parallel(1,1,0)$.
 }\label{fig_3D_band_kz}
\end{center}
\end{figure}
In the above simplified model, the band touching points are aligned on a single energy.
However, the nodal-line in the realistic electronic states has the energy dispersion as shown in Fig.\ \ref{Nodal-line} (b), so that the 2D Dirac cone is either electron doped or hole doped depending on $k_z$.
The low frequency excitation is suppressed in a partially doped Dirac cone\cite{Ando2002} and then the dynamical conductivity reduces in the low frequency region.
This explains the upslope tendency of the dynamical conductivity in $\omega<350$ meV in small broadening parameter $\hbar/\tau$ [Fig.\ \ref{optical_conductivity}(a)]

Finally, we analyze the suppression of $\sigma_1(\omega)$ by tilt of polarizing direction to the $c$ axis.
Our simplified model can also describe the 2D band structure in the 2D momentum space parallel to $k_z$ along a vertical nodal line in Fig.\ \ref{Nodal-line}(a).
In Fig.\ \ref{fig_3D_band_kz}, we plot the band structure in such a 2D momentum space parallel to $(k_x,k_y)=(1,1)$ which is represented by $(k_1+\pi/(5a),k_1-\pi/(5a),k_z)$ in the 3D reciprocal space.
This band structure shows that the velocity along the $z$-axis is much smaller than that in the $k_1$ direction, and thus the dynamical conductivity is strongly suppressed when the polarizing direction is parallel to the $z$-axis as shown in Fig.\ \ref{optical_conductivity}(a).

\section{Conclusion}\label{Sec_conclusion}
In conclusion, we investigate the dynamical conductivity of ZrSiS by a multi-orbital tight-binding model based on the first-principles band calculation.
We analyzed the polarization dependence and the frequency dependence in connection with the nodal-line structure.
In particular, we found that the dynamical conductivity in the clean limit is actually not frequency-independent unlike the ideal Dirac model, due to the significant deviation from the simple linear dispersion in the low-energy band structure.
We introduce a simple $2\times2$ model , which well describes the qualitative feature of the dynamical conductivity of ZrSiS beyond the simple Dirac model. 

\begin{acknowledgments}
M.K. and T.H. acknowledge the support of JSPS KAKENHI Grants No. JP25107005, No. JP25107001, and No. JP17K05496.
\end{acknowledgments}

\bibliography{ZrSiS}
\end{document}